\documentclass{elsart41}
\usepackage{graphics}
\usepackage{graphicx}
\usepackage{amssymb}
\begin{document}

\begin{frontmatter}



\title{Orbital order and spin--orbit coupling in BaVS$_3$}
%

\author{K. Radn\'oczi} and 
\author{P. Fazekas\corauthref{Name2}}\ead{pf@szfki.hu}
\address{Research Institute for Solid State Physics and Optics, Budapest, P.O.B. 49, H-1525 Hungary}  \corauth[Name2]{Corresponding author. fax: (36)(1) 3922218}

\begin{abstract}
The correlated $3d$ sulphide BaVS$_3$ undergoes a sequence of three symmetry breaking transitions which are reflected in the temperature dependence of the magnetic susceptibility, and its anisotropy. We introduce a microscopic model based on the coexistence of wide band $a_{1g}$ and localized $e_g$ $d$-electrons, and give the  classification of the order parameters under the double space group and time reversal symmetries.  Allowing for the relativistic spin--orbit coupling, the $d$-shell multipoles acquire a mixed spin--orbital character. It follows that  orbital ordering is accompanied by a change in the susceptibility anisotropy. 

\end{abstract}

\end{frontmatter}


BaVS$_3$ can be regarded as a $d$-electron system with hidden order, or perhaps with several hidden orders. Its metal--insulator transition (MIT) has been known for over 25 years  \cite{Massenet}, and it was long supposed to be an example of a Mott transition without symmetry breaking. No magnetic ordering accompanies the MIT and it was not discovered until 2002 that there is a structural change: the unit cell is doubled along the trigonal  axis \cite{inami}. The structural order parameter could be used to construct a Landau theory, and this would be sufficient to account for the fact that the MIT remains a second order transition in a wide range of pressure \cite{kezs}.  Nevertheless, we would like to consider BaVS$_3$ as a correlated electron system, and find electronic (as well as structural) order parameters for its phases. A systematic search for order parameters is based on symmetry classification.
 
The high temperature symmetry group of BaVS$_3$ is  $G_s=P6_3mmc{\otimes}T_t$ where $T_t$ stands for time reversal. The unit cell contains two V sites along the $C_3$ axis, so the cell doubling mentioned previously means that the $T<T_{\rm MIT}$ unit cell contains four V atoms \cite{inami,pouget}.   

The symmetry is lowered from $G_s$ by a sequence of three ordering transitions:  a hexagonal-to-orthorombic transition at $T_s=250$K, the MIT at $T_{\rm MIT}=69$K (breaking translational symmetry) and finally, time reversal invariance is broken at $T_X=30$K. Certainly there is an antiferromagnetic ordering of the magnetic dipoles but it is  probably accompanied  by the ordering of composite spin--orbital octupoles. 
 
BaVS$_3$ is a $3d^1$ system. In the high-$T$  hexagonal phase, the $t_{2g}$ level is split into a $a_{1g}$ singlet, and an $e_g$ doublet of orbitals $|a\rangle$ and $|b\rangle$. It is a long-standing problem \cite{Massenet,Mihaly2000} whether only $e_g$ states are occupied, or both $a_{1g}$ and $e_g$ levels are important. The current understanding emerging from ARPES experiments \cite{mitrovic} and DMFT calculations \cite{georges} is that $a_{1g}$ and $e_g$ electrons are present in roughly equal proportions. The $a_{1g}$ states form a wide band, while $e_g$ electrons carry localized (dipole, quadrupole, and arguably octupole) moments. 

The fact that the orbital degrees of freedom of the $e_g$ electrons play an important role is shown by the temperature dependence of the anisotropy of the magnetic susceptibility \cite{Mihaly2000}. The source of anisotropy is that both $t_{2g}$ and trigonal $e_g$ electrons have unquenched orbital moment, and thus there is a first-order effect from the relativistic spin--orbit coupling (SOC). For vanadium compounds, the SOC is substantial ($\lambda\approx 0.1$eV). Susceptibility  anisotropy measured by $A=(\chi_c-\chi_a)/\chi_c$  begins to deviate from 0 at about 250K, reaches -.07 at 70K, and +0.2 at low $T$ ($c$ is the trigonal axis, $a$ one of the orthorombic axes). The transitions at $T_{\rm MIT}$ and $T_X$ are marked by cusp-like anomalies of $A$.  
Published results allow a comparison between $\chi_c$ and $\chi_a$ only, but we predict further anisotropy within the $a$--$b$ plane.

  The standard order parameter of the 250K transition is purely structural: it is the change in the $c/a$ ratio \cite{say}. Lifting the degeneracy of the $e_g$ level causes orbital polarization \cite{sanna} which we measure by the expectation value of the pseudospin $\tau_z=(1/2)(\sum_S |aS\rangle \langle aS| - |bS\rangle \langle bS| )$, where $S$ is the spin.  The transverse components $\tau_x$, $\tau_y$ of the orbital pseudospin are analogously defined. The time-reversal-odd component is $\tau_y=(1/2)L_z$, the orbital momentum along the trigonal axis.

The evidence from the susceptibility anisotropy gives sufficient reason to assert that the orthorombic splitting must be modest and therefore the orbital degree of freedom $\tau$ remains important at all $T$'s.  

We performed a symmetry analysis and preliminary mean field calculations for two cases: 1) considering a single V site as a unit, and 2) for the  two-V-atom unit cell of the high-$T$ structure of BaVS$_3$. 
The details will be given elsewhere \cite{3}; here we restrict ourselves to simple illustrative arguments.

A single $t_{2g}$ electron in external fields can be described by
\begin{equation}
H =  \lambda{\bf S}{\cdot}{\b F} + \Delta (1-L_z^2) - q O_{xz}  -{\bf H}{\cdot}({\bf L}+2{\bf S})\label{eq:1}
\end{equation}
where the first term is the SOC, the second the trigonal field, the third 
is a quadrupolar mean field corresponding to the ordering of
\begin{equation}
O_{xz} = \frac{1}{2}(L_x L_z + L_zL_x)\label{eq:2}
\end{equation}
and the last is the Zeeman term. We arbitrarily chose one member of quadrupolar order parameter doublet $\{O_{xz},O_{yz}\}$. We find that 
$O_{xz}$ induces the mixed spin--orbital quadrupole
\begin{equation}
O_{xz}^{\prime} = \frac{1}{2}(L_x S_z + L_zS_x)\label{eq:3}
\end{equation}
The magnetic field effects are:
\begin{enumerate}
\item{for ${\bf H}\parallel (0,0,1)$, in addition to $S_z$ and $L_z$, also $S_x$ and $L_x$ are induced.}
\item{ for ${\bf H}\parallel (1,0,0)$, in addition to $S_x$ and $L_x$, also $S_z$ and $L_z$ are induced.}
\item{${\bf H}\parallel (0,1,0)$ induces only $S_y$ and $L_y$.}
\end{enumerate}
Analogous results are found for other kinds of orbital order. In principle, we can hope to identify the type of orbital order from finding  which non-diagonal elements of the susceptibility tensor have non-zero values. 

In contrast to the cubic $e_g$, the trigonal $e_g$ doublet carries both dipoles and octupoles. The argument given above is equivalent to stating that the spin--orbital octupole $\tau_xS_x-\tau_zS_y$ has the same symmetry as $S_z$ . This may be relevant for the description of the nature of ferromagnetism of BaVSe$_3$ \cite{yama}.  

Going over to the case of the two-atomic unit cell, we have to worry about the spatial distribution of $e_g$ and $a_{1g}$ electrons. 
To a good approximation, there is one localized $e_g$ electron per two V atoms, i.e., one $e_g$ electron in each high-$T$ unit cell containing the sites V1 and V2. We developed a Kondo-necklace-type \cite{doniach} spin-pseudospin model in which all ordering degrees of freedom are ascribed to $e_g$ electrons \cite{3}. In addition to the spin ($S$) and orbital pseudospin ($\tau$) degrees of freedom, we have a second pseudospin $\eta$ describing the position of the $e_g$ 
electron in the unit cell ($\eta_z=1/2$ for V1, $\eta_z=-1/2$ for V2). The 8-dimensional local Hilbert space supports $8^2-1= 63$ local order parameters which we classified under the symmetry $G_s=P6_3mmc{\otimes}T_t$.

We carried out mean field calculations in the $S$--$\tau$--$\eta$ 
model, assuming a sequence of ferro-$\tau$ and antiferro-$\eta$ ordering \cite{3}. Alternating order of the $\eta_z$ pseudospins amounts to the appearance of a four-V-atom supercell with the 
$a_{1g}$--$e_g$--$e_g$--$a_{1g}$ character. Superexchange gives singlet binding at the $e_g$--$e_g$ centers of the tetramers, explaining  the observed spin gap \cite{kezs}. Spin correlations are drastically reduced in the  para-$\eta$ phase, though orbital order persists.

\section*{Acknowledgement}
We acknowledge support by the Hungarian National Grant T 038162.
P.F. is grateful to L. Forr\'o, A. Georges, T. Kobayashi, G. Mih\'aly, H. Nakamura and K. Penc for enlightening discussions.

\end{document}